\documentclass[12pt]{article}

\usepackage{cite}
\usepackage{amsmath,amssymb,amsfonts}
\usepackage{algorithmic}
\usepackage{graphicx}
\usepackage{textcomp}
\usepackage{xcolor}
\usepackage{hyperref}

\begin{document}

\title{From Niche to Mainstream: Community Size and Engagement in Social Media Conversations}

\author{
    Jacopo Nudo$^{1}$, Matteo Cinelli$^{1}$, Andrea Baronchelli$^{2}$, \\
    and Walter Quattrociocchi$^{1}$
}

\date{}

\maketitle

\begin{center}
    $^{1}$ Center of Data Science and Complexity for Society, Department of Computer Science, Sapienza University of Rome, Rome, Italy \\
    $^{2}$ Department of Mathematics, City St George's, University of London, and The Alan Turing Institute, London, UK \\
\end{center}



\maketitle

\begin{abstract}
The architecture of public discourse has been profoundly reshaped by social media platforms, which mediate interactions at an unprecedented scale and complexity. This study analyzes user behavior across six platforms over 33 years, exploring how the size of conversations and communities influences dialogue dynamics. Our findings reveal that smaller platforms foster richer, more sustained interactions, while larger platforms drive broader but shorter participation. Moreover, we observe that the propensity for users to re-engage in a conversation decreases as community size grows, with niche environments as a notable exception, where participation remains robust. These findings show an interdependence between platform architecture, user engagement, and community dynamics, shedding light on how digital ecosystems shape the structure and quality of public discourse.
\end{abstract}

\section{Introduction}
Social media platforms have transformed online participation, becoming integral to daily life as primary sources of information, entertainment, and communication\cite{tucker2018social, gentzkow2011ideological, aichner2021twenty}.
Although these platforms provide unprecedented opportunities for connectivity and information sharing, their integration with complex social dynamics has raised significant concerns about their social impact. 
A major challenge in studying how platforms influence human behavior is the lack of accessible data \cite{roozenbeek2022democratize}, and even when researchers obtain data through special agreements with companies such as Meta, it may still be insufficient to distinguish inherent behaviors from the effects of platform design \cite{gonzalez2023asymmetric,guess2023social,guess2023reshares,nyhan2023like}.
This difficulty arises because the data, deeply embedded in platform interactions, complicate separating intrinsic human behavior from the influences exerted by the platform’s design and algorithms.
Research has delved deeply into issues such as polarization, misinformation, and antisocial behaviors in digital spaces \cite{del2016spreading,bail2018exposure,lupu2023offline,lazer2018science,cinelli2021dynamics}, highlighting the intricate and multifaceted effects of social networks on public discourse. 

In this context, a key area of research focuses on identifying differences in dialogue dynamics across platforms, exploring whether these are shaped by algorithmic or design choices, or by community-driven factors\cite{Avalle2024,waller2021quantifying,di2024patterns}.
The attention a piece of content gets on a platform helps measure its reach, impact, and spread. When users comment repeatedly on a post, it shows they are deeply engaged and willing to share their thoughts\cite{budak2024misunderstanding, xiao2022does}.

This study compares public conversations, combining a long-term perspective with a multi-platform dataset to examine their dynamics across different digital environments. We define an interaction as the comments a user posts within a thread. To measure user engagement—and, by extension, their attention—we analyze the number of comments left in a thread and the probability of re-entering the conversation after posting, following the methods outlined in \cite{backstrom2013characterizing}.
We investigate how these aspects are influenced by the platform where the conversation takes place and by size of the context, defined either as the number of people commenting in a thread or as the number of people that belong to a community or a page where the conversation is hosted.

Looking at the probability re-entry in a thread, that is similar to the probability of leaving more than one comment in a conversation, we notice its sensitiveness to size of the context, analogous to quorum sensing in bacteria, where population density governs collective behaviors through the exchange of signaling molecules \cite{miller2001quorum}; much like how the behavior of individual particles in a system often depends critically on the size of the system itself, as observed in phenomena like finite-size effects in phase transitions and critical phenomena, where collective behavior dominates individual dynamics \cite{stanley1971phase}.

So the main research questions are: 
\begin{description}
\item[{\bf RQ1}]Do users maintain consistent participation and re-entry patterns across various platforms?
\item[{\bf RQ2}] How does the size of the crowd participating in a conversation affect the probability of user re-entry?
\item[{\bf RQ3}] How does the outreach size of the community hosting the conversation impact the probability of user re-entry?
\end{description}

Using data from\cite{Avalle2024}, we systematically compare threads (or conversations) linked to posts on Usenet, Gab, Reddit, Voat, Twitter, and Facebook to answer these questions. The conversations are from a wide range of topics (vaccines, conspiracies, politics, news, Brexit, climate change) and cover 33 years.
This research provides actionable insights for platform designers and policymakers seeking to foster healthier online conversations.

\section{Related works}

The academic literature has explored various aspects of user behavior in online communities. 

Studies such as \cite{backstrom2013characterizing} analyzed the likelihood of users re-entering a conversation after leaving a comment, using specific participation patterns as predictors.

Researchers studied the structure of the conversation on Reddit, intended as a tree graph, and analyzed the local and global features, like post sentiment and Subreddit popularity\cite{yu2024characterizing}. While \cite{saveski2021structure, etta2024topology} investigated how the structure is connected to the toxicity level of the comments produced. Additionally, \cite{xiao2022does} suggests that persuasiveness increases as comments become more deeply nested in a thread structure.

Works like \cite{di2024users} analyze how users explore diverse options before settling into specific communities, highlighting their persistence within them and their consistency in discussing similar topics despite volatile engagement across different communities.
Studies such as \cite{hamilton2017loyalty} and \cite{mainwaring2017turnover} analyze loyalty and turnover among user or consumer in groups or subscription programs, looking for factors that influence it. 

In \cite{panek2018effects}, researchers demonstrated how variability in user activity changes significantly with the size of the community on Reddit, suggesting implications for platform design. Relatedly, literature on the attention economy investigated human attention \cite{davenport2001attention,falkinger2008limited,sangiorgio2024followers} as limited resource that platforms and communities compete to capture and retain.

A comprehensive multi-platform analysis of vocabulary usage is provided by \cite{di2024patterns}, exploring linguistic patterns and variations across various social media platforms. Finally, recent studies such as \cite{Avalle2024} have conducted multi-platform analysis to identify consistent patterns in toxic discourse dynamics, highlighting persistent patterns in toxic behavior across various platforms.
Despite the limited material, these studies provide a solid starting point for analyzing attention dynamics, offering a comprehensive overview of user behavior in online communities that can be used a solid background for further analysis.

\section{Materials and methods}
This section outlines the data sources for all analyzed social media platforms, detailing the data acquisition process, the preprocessing steps, and the measures employed in the analysis.

\subsection{Data Collection}

{\bf Usenet:} Usenet is one of the oldest communication networks on the internet, established in 1980. It is organized into a decentralized system of newsgroups, each dedicated to a specific topic or theme. Users post messages or “articles” to these newsgroups, creating public threads for others to view, read, and reply to. Due to its unmoderated and decentralized nature, Usenet is known for hosting a variety of niche discussions, some of which predate modern social networks. We collected data for the Usenet discussion system by querying
the Usenet Archive (https://archive.org/details/usenet?tab=about). We
selected a list of topics considered adequate to contain a large, broad
and heterogeneous number of discussions involving active and populated newsgroups. As a result, we selected conspiracy,
politics, news, and talk as topic candidates for our analysis. 

{\bf Facebook:} Facebook is a widely used social networking platform launched in 2004 that enables users to create personal profiles, add friends, and join groups organized by interests. Facebook centers around individual profiles and direct relationships, facilitating the sharing of posts, images, and comments among friends, as well as reactions such as likes. Users interact within groups or public pages to discuss topics of interest, share information, or engage with communities. We use datasets from previous studies that covered discussions about news \cite{schmidt2017anatomy} gathered from a list of pages by the Europe Media Monitor that reported the news in English.

{\bf Twitter:} Twitter (now rebranded X) is a social networking service where users post and interact through short messages known as "tweets." Users can follow others, retweet, reply, like, and engage in discussions around hashtags or trending topics. Twitter's timeline focuses on user-generated content organized around individual user accounts, interactions, and hashtags rather than groups or communities. We used a list of datasets from previous studies that includes
discussions about vaccines \cite{valensise2021lack}, climate change \cite{falkenberg2022growing} and news \cite{quattrociocchi2022reliability} topics.

{\bf Gab:} Gab is a social media platform launched in 2016, marketed as a “free speech” alternative to more mainstream networks. It is organized similarly to Twitter, with users following one another and posting status updates, comments, and reacting to other users' posts. Gab has become popular among users banned or discontent with other platforms due to content moderation policies\cite{zannettou2018gab}. We collected data from the Pushshift.io archive (https://files.pushshift.io/gab/) concerning public discussions.

{\bf Reddit:} Reddit is a social content aggregation website, organized in communities constructed around specific topics, named subreddits. Each user has an account corresponding to a user name used to post submissions or to comment on other submissions and other comments. In addition, users can also upvote or downvote a submission in order to show their appreciation or criticism for it. Differently from other social media, Reddit’s homepage is organized around subreddits and not on user-to-user relationships. Therefore, subreddits chosen by users are likely to represent their preferred topics and the main source of information consumed on the website. We collected public data from Reddit using the Pushift collection \cite{baumgartner2020pushshift}.

{\bf Voat:} Voat.co was a news aggregator website similar in structure to Reddit, operating until December 25, 2020. It gained attention as a migration hub for users banned from Reddit, hosting discussions in specialized communities called "subverses." Users could subscribe to subverses of interest and interact with content through comments, upvotes, and downvotes. Given the platform’s alignment with certain controversial topics, Voat’s user community became a focal point for discussions that were censored on other platforms. For this study, we utilized a dataset collected from \cite{mekacher2022can}, which captures a snapshot of interactions on Voat prior to its shutdown.

\subsection{Preprocessing}

According to the social media description provided in the previous section, each platform provides post-related discussion in the form of comments threads.
For each platform, such comment threads were extracted and for each comment, the \textit{timestamp} of the publication, the \textit{post\_id} (i.e the thread identifier), and the \textit{user\_id} were obtained. If the \textit{user\_id} was not available due to anonymization, the comment was not included.
Table \ref{tab:data_breakdown} shows a data breakdown.


\begin{table}[!ht]
    \centering
    \begin{tabular}{|c|c|c|c|}
    \hline
         & Users & Comments & Threads  \\
         \hline
        Gab & 166,833 & 12,509,891 & 3,764,443 \\
        \hline
        Reddit & 394,733 & 1,798,628 & 808,016 \\
        \hline
        Twitter & 13,620,442 & 22,337,801 & 325,451 \\
        \hline
        Usenet & 212,259 & 3,566,773 & 682,362 \\
        \hline
        Voat & 153,255 & 3,454,791 & 413,854 \\
        \hline
        Facebook & 17,700,372 & 39,732,512 & 1,000,000 \\
        \hline
    \end{tabular}
        \vskip 5mm
    \begin{tabular}{|c|c|}
    \hline
         Platform & Time Range  \\
         \hline
        Gab & Oct 2016 to Oct 2018 \\
        \hline
        Reddit & Jan 2018 to Dec 2022 \\
        \hline
        Twitter & Jan 2010 to Jan 2023 \\
        \hline
        Usenet & Feb 1989 to Jun 2013 \\
        \hline
        Voat & Nov 2013 to Dec 2020 \\
        \hline
        Facebook & Dec 2009 to Aug 2016 \\
        \hline
    \end{tabular}
        \vskip 5mm
    \caption{Data breakdown.}
    \label{tab:data_breakdown}
\end{table}

\subsection{Measure of Users Participation}
\label{Participation}

In order to study how the number of unique participants varies with conversation length, we provide a measure of users' tendency to participate in comments threads. Following the method used in\cite{backstrom2013characterizing}, we construct a two-dimensional \textit{density matrix} that describes the distribution of user participation across conversations of different lengths.
Given a list of comments, where \( u \) is the group of users who produced the comments, let \( d = |u| \) represent the number of unique users,  and let \( k \) denote the number of comments in the list.
For each value \( k \), where \( k \) ranges from 0 to 200, we select all threads \( t_{k,p} \in T_p \), where \( T_p \) represents the set of threads on platform \( p \), that contain at least \( k \) comments. Again for each value of \( k \), we select among $t_{k,p}$ just the first $k$ comments for each thread, and we calculate the relative frequency of threads having $d$ users where $d \in [1,k]$. These relative frequencies are contained in a density matrix \( D \), where each cell \( c_{d,k} \) holds the value \( P(d|k) \), the probability that within the first \( k \) comments there are \( d \) unique users. Note that column-wise the matrix \( D \) contains discrete probability and thus each column sums up to one. The patterns that may emerge from the density matrix highlight different participation dynamics. For example, a distribution near the diagonal indicates a “broad-spectrum conversation,” where new users frequently contribute a single comment each. Conversely, density concentrated towards the bottom of the matrix suggests a “concentrated participation,” indicating that a few users produce most of the comments. The second case reflects intense discussions among a small group of interlocutors, that stand for longer attention and reading time.

\subsection{Measure of User Re-entry}
\label{Measure}
To assess the activity of a user posting multiple comments and engaging in dialogue, we choose a measure that evaluates the concentration associated with the distribution of the number of users making $k$ comments within a thread. This measure is based on the second and fourth moments of the distribution. 
Defined $k$ as the interaction length, we compute the following metric:

\[
L = \frac{\left( \sum_{k}\phi_{k}^{2} \right)^{2}}{\sum_{k}\phi_{k}^{4}}
\]

Where \(\phi_{k}\) represents the probability that a user writes \(k\) comments under the same thread. This measure, \(L\), is intended to quantify the degree of concentration in the distribution of comments, and the position of the peak \cite{nagel1984phonon}. A higher value of \(L\) indicates that users tend to leave about \(L\) comments on the same thread, while a low value, close to 1, indicates a general habit of leaving just one comment and leaving the conversation. This measure provides insight into user engagement within discussions, helping us to understand how participation varies among users and whether certain conversations present users with different levels of commitment. It also provides a mathematical basis for comparing the distribution of comments across different threads or posts. Similar techniques based on higher moments of distributions have been used in the study of social network behaviors \cite{quattrociocchi2014opinion}.

\begin{figure}[!ht]
\centering
\includegraphics[width=\linewidth]{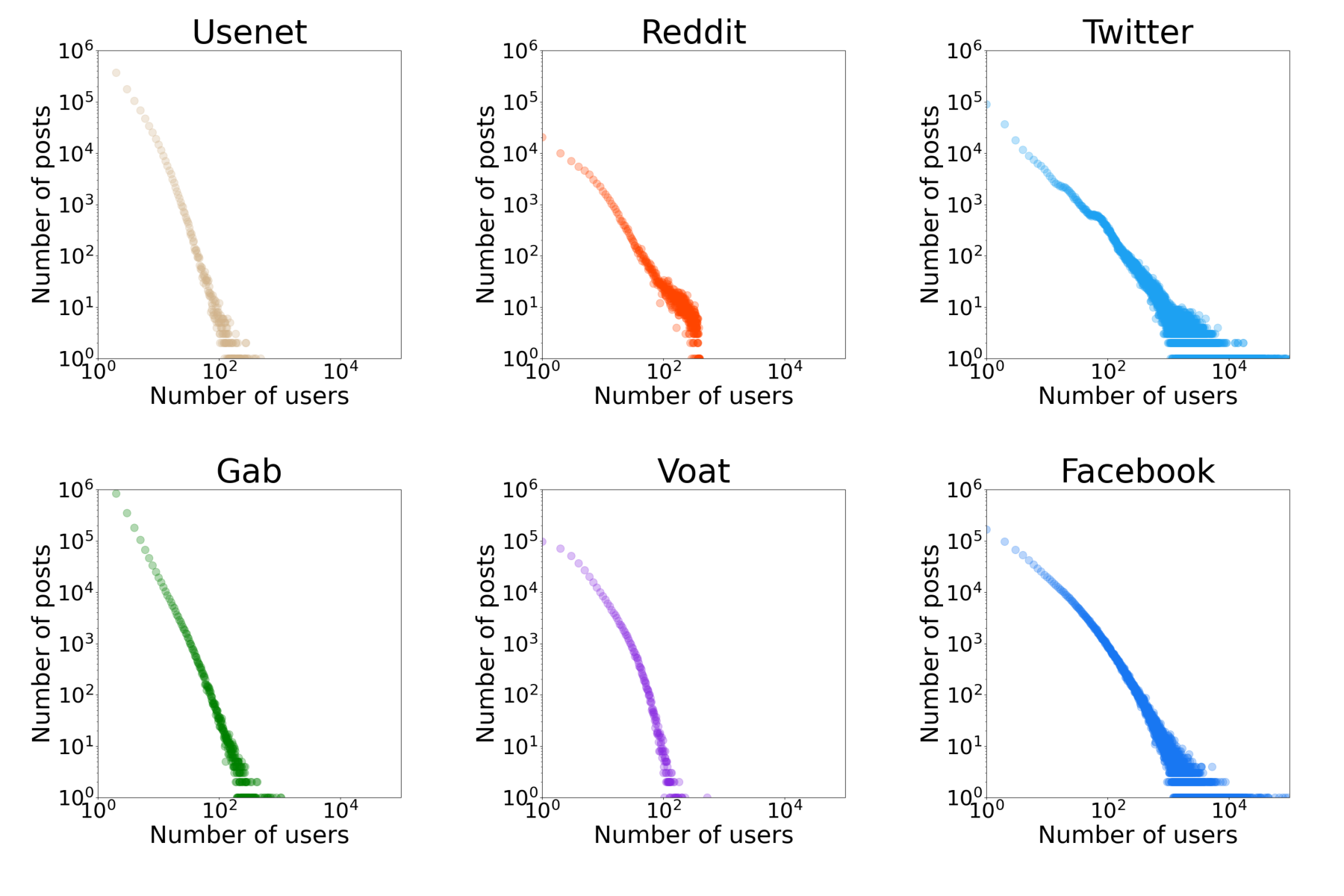}
\caption{Log-log distribution of the number of unique users commenting under a post, across different platforms. The y-axis represent the absolute frequency of posts that have this amount of unique user. Note the heavy-tailed distributions typical of power-law behaviors.}
\label{fig:1}
\end{figure}

\subsection{Measures of Size}

To investigate how individual user engagement, measured by re-entry probability, is influenced by the size of the social context in which they are embedded, we define size using two distinct metrics: (i) the number of users who have contributed to a comment thread, referred to as the \textit{crowd}, and (ii) the number of individuals visiting the page or group hosting the conversation at a given time, termed \textit{outreach}.

\label{crowd}
The first way to define the size is using the number of unique users who have participated in a comment thread (crowd).
Given $u$ the list of all the users that commented in a specific thread, the size of the crowd is calculated as \( d = |u| \) . 
So for each interaction, done by a user in a thread, we can determine the size of the crowd in the thread. Then we can sort by $d$, and divide all interactions in bins, for each bin calculate the localization index (\( L \)), as explained in \ref{Measure} on the distribution of the interaction length \(\phi\).

\label{Outreach}
The second way to define the size is as the number of users involved in the community of a page or a group that host the conversation, defining it as the outreach.
Following \cite{sangiorgio2024followers}, we start counting the number of users commenting on a certain week.
Given \(p\) a page (or a group) and \(t\) a specific date, we define \(O_{p}(t)\) as the outreach of that page, calculated as the number of users who have commented during that week, smoothed by a rolling average over a time window of 12 weeks.
For each interaction done on each page at time $t$, we can determine the number of active users on that page \(p\) at that time using \( O_{p}(t) \). Defining an interaction as the set of comments written by the same user under a thread, we sort all the interactions based on the outreach (\( O\)), and divide them into bins of 1000 observations. Finally, for each bin, we calculate the localization index (\( L \)), as explained in \ref{Measure}, to determine whether the typical number of comments, or the re-entry probability of a user, are connected to the level of outreach of the page commented .

\section{Results and Discussions}
\subsection{Distribution of Number of Users Across Platforms}

Figure~\ref{fig:1} displays the distribution of the number of unique users per post on the 6 social media platforms: Gab, Reddit, Twitter, Usenet, Voat, and Facebook. The x-axis represents the number of unique users, and the y-axis represents the number of posts being commented by $x$ users, both plotted on logarithmic scales.

Across all platforms, the distributions exhibit a power-law-like behavior (See Supplementary Information), characterized by a heavy-tailed shape. This suggests that while most posts receive comments from only a small number of users, a small proportion of posts achieve significantly higher engagement. Such distributions are typical in online platforms \cite{Avalle2024}, where content virality and user behavior follow non-uniform patterns \cite{barabasi2001physics}.

Platforms within the same columns in the figure show similar interaction patterns.
Notably, Twitter and Facebook share broader distributions with extended tails that reflect the occurrence of highly viral posts with significant user interactions.

\subsection{Analyzing User Participation}

\begin{figure}[!ht]
\centering
\includegraphics[width = 1\linewidth]{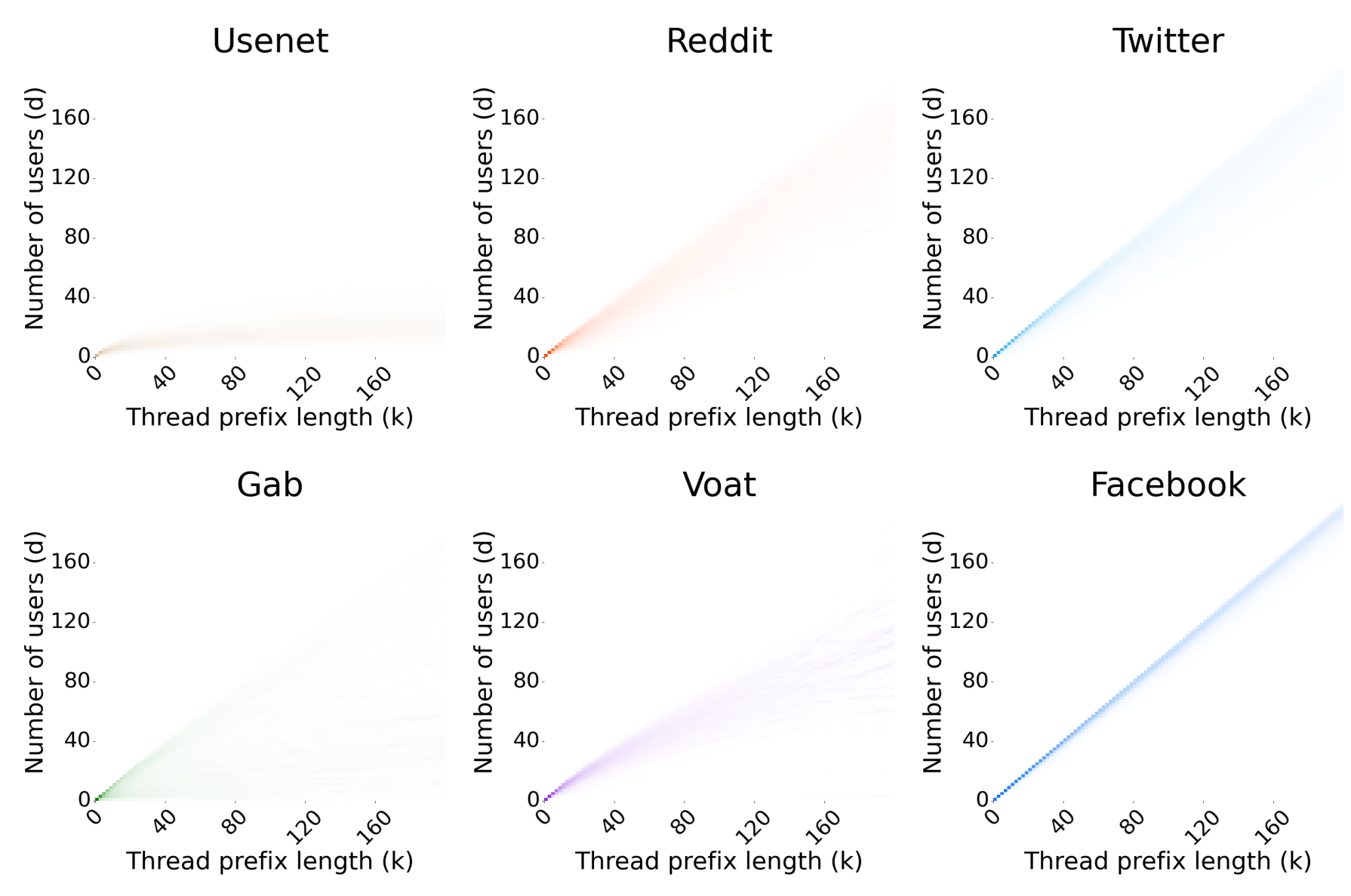}
\caption{Density matrices illustrating the distribution of a number of unique users (\( d \)), given thread prefix length (\( k \), number of comments), across different platforms.   The diagonal trend on platforms like Facebook and Twitter indicates a proportional increase in unique user engagement with thread length, while the lower concentration on platforms suggests a smaller core group driving most of the activity. Color intensity reflects the probability densities (conditioned by column), highlighting platform-specific interaction patterns.}
\label{fig:2}
\end{figure}

To address {\bf RQ1}, we investigate the heterogeneity of participation across a set of comments and the likelihood of re-entering a conversation on various platforms. 
The density matrices for user participation on Reddit, Twitter, Usenet, Voat, Facebook, and Gab illustrate how conversation length correlates with the number of unique users, represented by the probability distribution \( P(d \mid k) \), as shown in Figure \ref{fig:2}. Here, \( k \) denotes the prefix length applied to the number of comments in a thread, while \( d \) represents the number of unique users contributing, as explained in \ref{Participation}.

Facebook and Twitter exhibit "expansionary" dynamics, where the density trends along the diagonal. This pattern suggests that as the prefix \( k \) increases, a comparable number of unique users \( d \) engage. Such dynamics indicate broad participation and diverse user interaction within conversations.
This comparison underscores different engagement styles: broad, transient participation on platforms like Facebook and Twitter versus deeper, concentrated discussions on Reddit, Voat, Gab, and Usenet.

\begin{figure}[!ht]
\centering
\includegraphics[width = \linewidth]{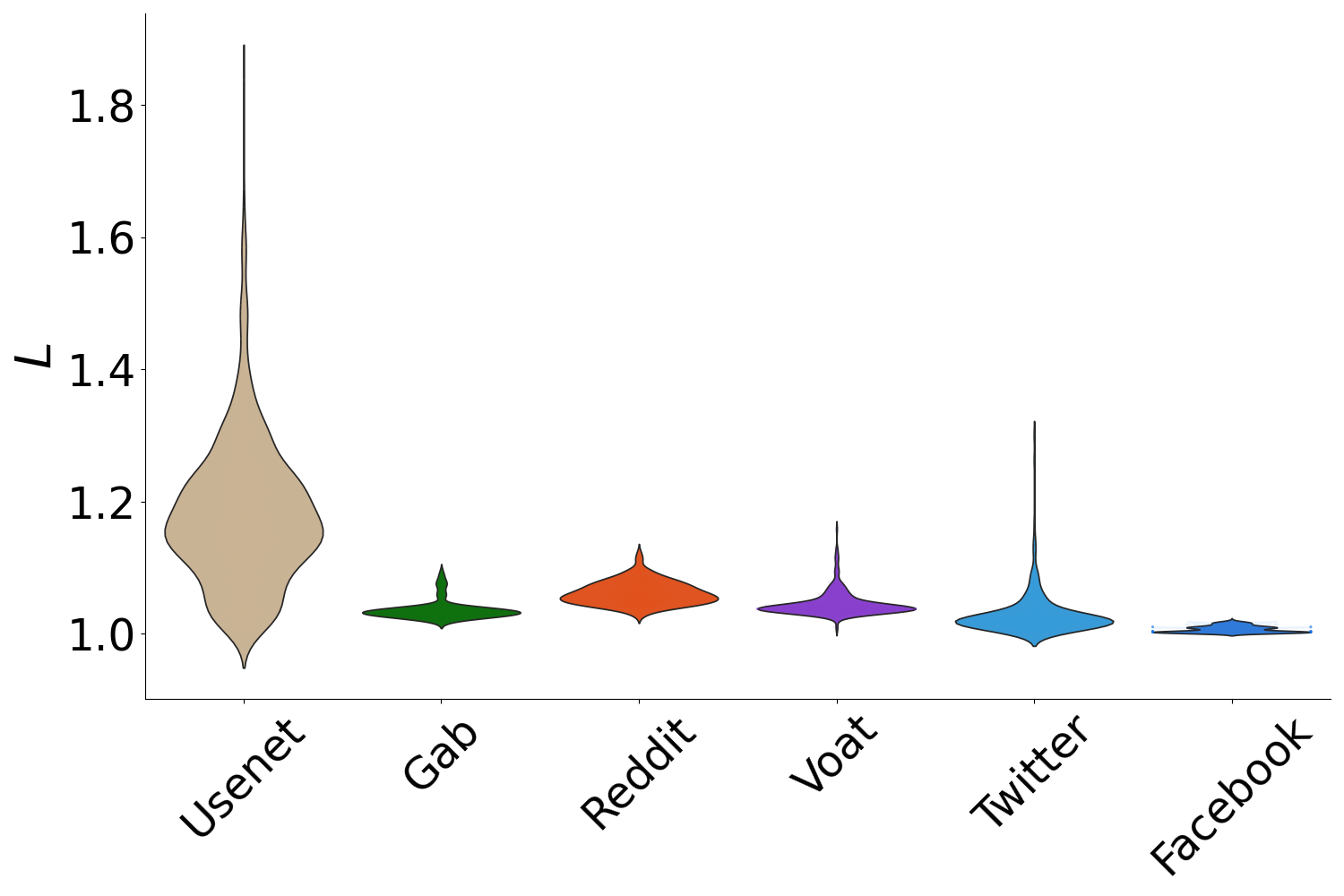}
\caption{Distribution of the localization parameter ($L$) across different platforms. 
The violin plot shows , using $L$, probabilities that an interaction (among a user and a post) is composed by a certain amount of comments, with each platform exhibiting distinct patterns. Localization values greater than 1 indicate a greater propensity to leave more than one comment in a conversation. While a value close to 1 stand for a distribution with less degrees of freedom.}
\label{fig:3}
\end{figure}

Thread-wise the re-entry patterns, measured using the number of comments by the same user under a post, is interpretable by applying the localization $L$ to the distribution of number of comments per user (calculated for splitting by week the datasets), as explained in \ref{Measure}. From Figure \ref{fig:3}, we note a certain degree of platform dependency in Localization distributions. Consistent with previous observations, Facebook exhibits the narrowest distribution ($L \simeq 1$), followed by Twitter. This observation is further supported by a Kruskal-Wallis test ($H = 1539.84$, $p < 0.001$), which indicates significant differences across platforms. A post-hoc Dunn test with Bonferroni correction revealed that most pairwise comparisons were significant, with exceptions observed between Gab and Voat ($p = 0.173$) and Gab and Twitter ($p = 0.311$).

Lower values of \(L\) can be interpreted as a reduced likelihood of leaving more than one comment under the same post. This measure appears to be distributed similarly across the entire analyzed time span; however, it seems to depend on the number of users participating in the conversation.

In order to obtain further evidence with respect to that provided by analyzing the index \(L\), we calculated the median of the probability distribution that a user-post interaction consists of a single comment for each platform. The medians are as follows: for Usenet, \(0.68\); for Gab, \(0.83\); for Reddit, \(0.78\); for Voat, \(0.82\); for Twitter, \(0.87\); and for Facebook, \(0.94\). (See Supplementary Information).

\subsection{User Re-Entry vs Crowd Size}

\begin{figure}[!ht]
\centering
\includegraphics[width = \linewidth]{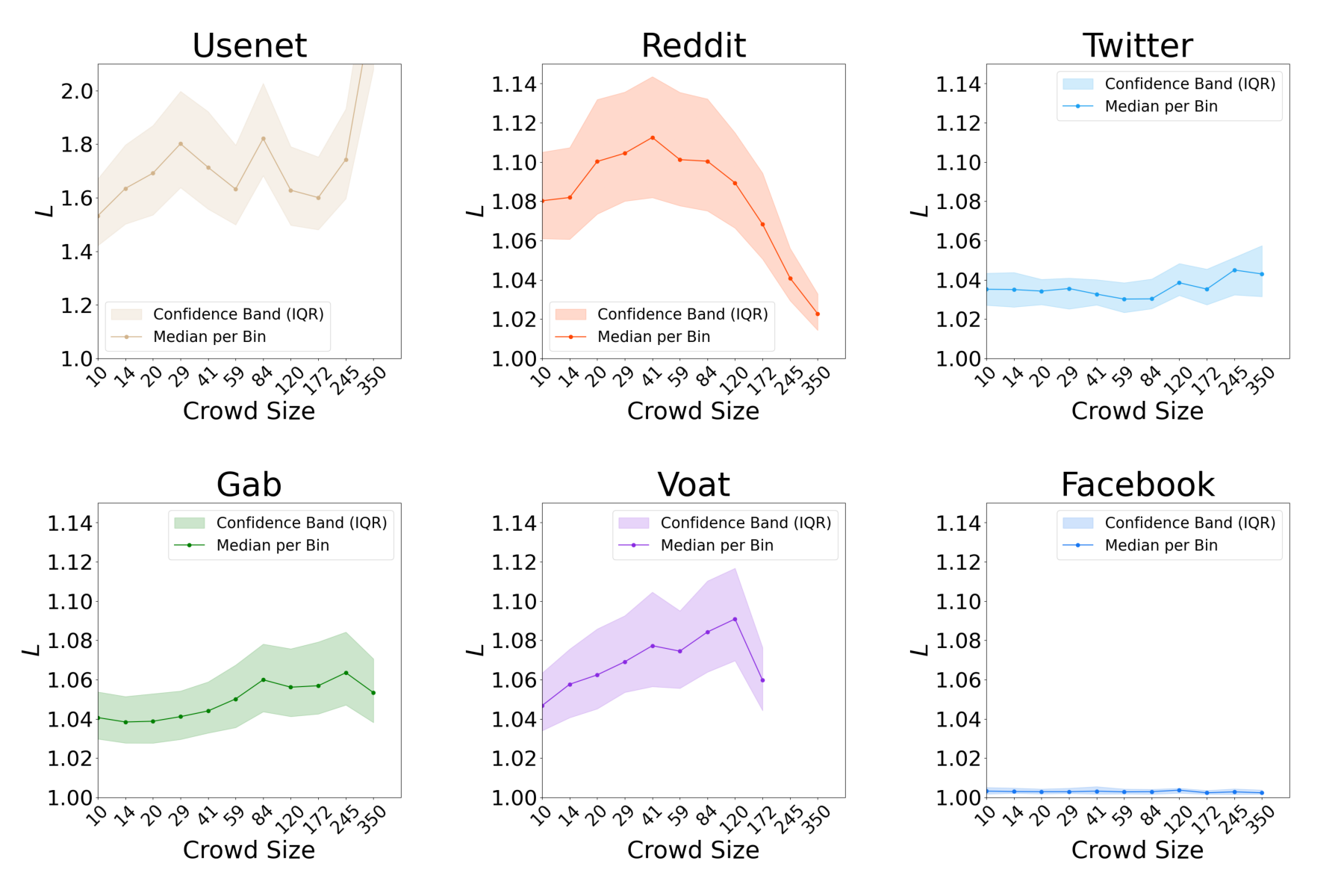}
\caption{Distribution of the localization parameter ($L$) across different platforms, controlling by the number of users involved in the thread (crowd size on x-axis). The number of users involved seems to influence the distribution of the number of comments posted by each user under the same post. Linearly for Usenet and Gab, with a saturation effect on Reddit and Voat, while it does not seem to impact giant mainstram platform such as Twitter and Facebook.}
\label{fig:4}
\end{figure}

In order to answer {\bf RQ2}, we define size of the crowd commenting under a thread, as explained in \ref{crowd}, in order to study its impact on re-entry probability. Once divided all the interactions, i.e. defined an interaction as the set of comments that a user writes under a thread, into bins with respect to the number of users involved (crowd size) we plotted the distribution of the localization ($L$) parameter for each bin, Figure \ref{fig:4}. For Gab and Usenet, it is possible to observe how, as the number of users in a conversation increases, the typical number of comments shifts away from localization values equal to one. While on Reddit and Voat, user engagement initially increases with the number of participants, peaking around 150 users. Beyond this threshold, additional participants reduce marginal engagement, likely due to conversational noise or decreased cohesion. 

Our findings reveal a bound in platforms like Reddit and Voat, where the likelihood of re-entering a conversation peaks around 150 participants. Surprisingly, this quantity aligns with the Dunbar Number, which represents the cognitive limit of stable social connections\cite{dunbar1998social,goncalves2011validation}. This boundary observed also in other domains such as news consumption\cite{cinelli2020selective}, app-usage\cite{de2019strategies}, and points of interests\cite{alessandretti2018evidence}, suggests that conversational dynamics in these environments are bounded not only by platform design but also by cognitive and social factors intrinsic to human behavior.
On smaller platforms like Usenet and Gab, we observe a linear increase in user engagement as the number of participants grows. Conversely, larger platforms such as Facebook and Twitter exhibit independence from community size, likely due to thread structures that promote shallow interactions.
These differences may stem from platform design choices that either constrain or encourage deeper dialogues. Smaller, niche platforms provide a focused environment, while mainstream platforms prioritize scalability, often at the expense of sustained user engagement.

\subsection{User Re-Entry vs Outreach Size}
In order to answer {\bf RQ3}, we investigate whether user propensity for dialogue is related to the size of the community associated to a page or a group (outreach size); we follow the methods described in \ref{Outreach}.
On each platform, we analyze user-post interactions and categorize each interaction in a bin of 1000 observations based on their outreach (\(O\)) level. To assess the impact of the average page outreach on bin's localization parameter (\(L\)), we perform a linear regression analysis. This approach allows us to examine how the distribution of re-entry varies as a function of increasing page outreach.

\begin{figure}[!ht]
\centering
\includegraphics[width = \linewidth]{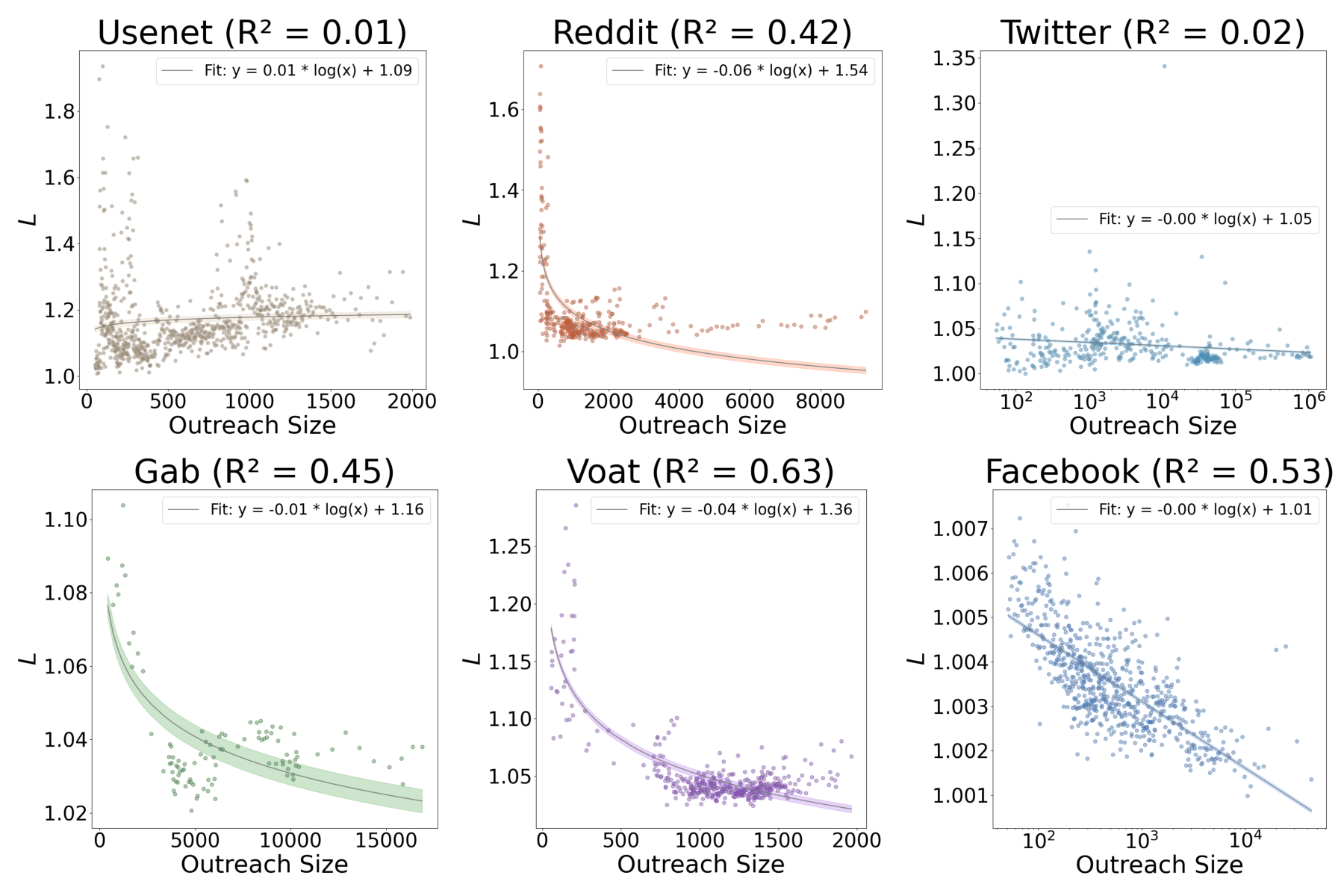}
\caption{Plot of the localization parameter $L$ (probability of re-entry on y-axis) changing the outreach level of the community (outreach size on x-axis), across different platforms.
Values on x-axis are represented in a Log scale for Facebook and Twitter.
Each interaction has been grouped in a bin, based on the level of outreach that the community (or the page) hosting the interaction had at the time. 
It show how, almost everywhere the probability of re-entry in a conversation decrease with the increase of the outreach size,  likely due to a progressively noisier conversation environment.}
\label{fig:5}
\end{figure}

As shown in Figure \ref{fig:5}, there appears to be a general trend across most platforms where the propensity re-entry in a conversation (i.e., to post multiple comments) is negatively correlated with the outreach size of the community that host the thread. This is due to an increasingly noisy and chaotic environment, where it becomes difficult to follow the thread of the conversation, leading to a more passive and distracted interaction.

This trend is confirmed by all platforms except Twitter and Usenet. On Twitter user behavior appears to be largely unaffected by the size of the outreach, with a general tendency to avoid extended dialogue. In the second case, data suggest that individuals are more likely to persist in conversations as the number of users increases and thus the stimulus.

\
\section{Conclusions} 
This study analyzes public conversations across six social media platforms over 33 years, uncovering significant differences in conversational dynamics and user engagement. The findings reveal how platform design, community size, and user behavior interact to shape dialogue in digital spaces, providing a nuanced understanding of the factors driving engagement.
Responding to RQ1, we find that the propensity for dialogue, measured as the number of re-entries in a conversation, is generally lower on larger, mainstream platforms. These environments favor broader but less sustained participation than smaller, niche platforms.
Addressing RQ2, our analysis shows that the propensity for dialogue correlates with the crowd size—the number of users involved in a conversation on smaller platforms. Initially, an increase in crowd size stimulates richer and more diverse interactions. However, conversations fragment and lose cohesion beyond a critical threshold, particularly on platforms like Reddit and Voat. This threshold aligns with cognitive limits, such as Dunbar's Number, highlighting the role of human constraints on social connections in shaping interaction quality within larger groups.
About RQ3, as platform environments scale, user engagement—measured by the likelihood of re-entering a conversation—declines. Across all platforms, larger communities and higher outreach correlate with reduced dialogue persistence. This trend is most pronounced on mainstream platforms like Facebook and Twitter, where content saturation and algorithmic amplification likely drive passive consumption and diminish individual involvement. In contrast, niche platforms maintain higher levels of engagement even as their community size increases, likely due to their focused and less algorithmically driven environments.
These findings underscore important implications for platform design and policy-making. By prioritizing scalability and virality, mainstream platforms may inadvertently undermine opportunities for meaningful dialogue, fostering fragmented interactions and passive use.
Future research explores how platform policies, user migration, and algorithm changes influence conversational dynamics. Longitudinal studies investigate the link between conversational fragmentation and the effectiveness of policies fostering constructive discourse.
In conclusion, public conversations on social media reflect a complex interplay between platform design, community size, and user behavior. Balancing scalability with meaningful interactions is key to fostering constructive dialogue and healthier online environments. Understanding these dynamics helps design better digital spaces for engaging discussions.

\bibliographystyle{unsrt}
\bibliography{main}

\end{document}